\documentclass[12pt]{iopart}

\usepackage{citesort}

\expandafter\let\csname equation*\endcsname=\relax
\expandafter\let\csname endequation*\endcsname=\relax
\usepackage{amsmath,amssymb,amsthm} % note amssymb will load amsfonts per Barbara Beeton also to avoid physics use
\usepackage{braket} % as suggested by marmot and this example answer https://tex.stackexchange.com/questions/316836/the-bra-ket-notation-in-latex

\usepackage{bbm}
\usepackage{graphicx} % Include figure files

\newcommand{\bsig}{\boldsymbol{\sigma}}
\newcommand{\bV}{{\bf V}}
\newcommand{\bv}{{\bf v}}
\newcommand{\bp}{{\bf p}}
\newcommand{\bR}{{\bf R}}
\newcommand{\bA}{{\bf A}}
\newcommand{\bB}{{\bf B}}
\newcommand{\bL}{{\bf L}}
\newcommand{\bk}{{\bf k}}
\newcommand{\brr}{{\bf r}}
\newcommand{\bmm}{{\bf m}}

\newcommand{\mel}[3]{\left\langle #1 \left| #2 \right| #3 \right\rangle}

\begin{document}

\title{Bulk-to-surface misorientation and the spin texture of topological insulators}

\author{N. Klier$^1$, F. Rost$^1$, R. Gupta$^2$, S. Sharma$^3$, O. Pankratov$^1$, S. Shallcross$^3$}
\address{1 Lehrstuhl f\"ur Theoretische Festk\"orperphysik, Staudtstr. 7-B2, 91058 Erlangen, Germany}
\address{2 H. H. Wills Physics Laboratory, University of Bristol,Tyndall Avenue, Bristol BS8 1TL, United Kingdom}
\address{3 Max-Born-Institute for Non-linear Optics and and Short Pulse Spectroscopy, Max-Born Strasse 2A, 12489 Berlin, Germany}

\date{\today}

\begin{abstract}
Weak topological insulators possess a symmetry related set of Dirac-Weyl cones in the surface Brillouin zone, implying misorientation between the principle axis   of the low energy manifold of the bulk and the surface normal. We show that this feature of weak topological insulators comes with a hidden richness of surface    spin textures, and that by misorientation a helical texture can become an unusual hyperbolic spin texture. We illustrate this effect by comparison of the $M$-     point and $\Gamma$-point Dirac-Weyl cones on the (111) surface of the crystalline topological insulator SnTe.
\end{abstract}

\maketitle

\section{Introduction}

Topological  insulators  (TI)  are  a  state of matter revealed only by the presence of a boundary and are, therefore, materials for which the bulk-boundary relationship is of profound importance\cite{yoi13}. The connection between the microscopic variables of the bulk insulator and the emergent effective degrees of freedom of the topological surface state (pseudospin and chirality), and how these two are entwined by surface potentials such as band bending, remains a fascinating question\cite{joz11,sou11,moo11,henk12,hai13,pan13,jak15}. Here we wish to address, in a realistic setting, the role played by a misorientation between the bulk manifold and the surface normal in the physics of the surface state. The experimental motivation for this is twofold: (i) in weak topological insulators bulk-boundary misorientation will be the generic situation and (ii) for SnTe-type topological insulators\cite{hsi12,liu13} the M points of the (111) surface allow access to a system in which the relationship between misorientation, band bending, and spin texture can be explored\cite{shi14}.

One of the most compact and insightful ways to explore the physics of the TI surface is through bulk continuum Hamiltonians, obtained for example by the $\bk.\bp$ method, combined with a suitable material-vacuum boundary condition. This approach has been extensively explored, and several different theoretical treatments given of the appropriate boundary conditions\cite{zha12}. In this work we
employ the supersymmetry theory of a band inversion interface, first considered in the context of IV-VI semi-conductor hetero-junctions by Volkov and Pankratov\cite{vol78,vol83,vol85,pan87,pan87a,pan89}. This treats the gap inversion and band bending on an equal footing and, importantly, the surface state is ``universal" in form, independent of the particular shape of the interface, requiring only sign change in the asymptotic values of the so-called ``super-potential", a function combining the band bending and gap functions. This approach to the interface problem we combine with a bulk Dirac equation derived from a tight-binding analysis of SnTe, providing the crucial gap edge wavefunctions required to incorporate the bulk microscopic degrees of freedom into the theory.

We  find  that misorientation between the bulk and surface coordinate systems results in a rich surface spin structure in which helical and hyperbolic spin textures can be tuned into each other by band bending. Crucial to the emergence of the hyperbolic texture is a rotation in the Kramers degenerate sub-space of the band gap functions induced by the misorientation. Whether a helical or hyperbolic spin texture is realized depends, as we show, sensitively on the microscopic physics of the material through the balance of spin-orbit and crystal field effects in the bulk. For the case of the M point  surface states of the (111) facet of SnTe, significant band bending (either downward or upward) results in a helical spin texture, with  intermediate downward band bending resulting in a hyperbolic spin texture.
 
\section{Model of SnTe bulk}

SnTe has a simple rocksalt crystal structure and at each of the 4 inequivalent L points of the fcc Brillouin zone the low energy band structure is well described by an effective Dirac equation

\begin{equation}
 H_D= \begin{pmatrix}
  \Delta_0 & \bsig\bv'\bp' \\
  \bsig\bv'\bp' & -\Delta_0
 \end{pmatrix}
 \label{eq1}
\end{equation}
with velocity tensor 

\begin{equation}
 \bv'=\begin{pmatrix}
       v_\perp & 0 & 0\\
       0 & v_\perp & 0\\
       0 & 0 & v_\parallel
      \end{pmatrix}
\end{equation}
which is diagonal in the local frame in which the $z'$ axis is directed along the L point vector $\bL=\frac{\pi}{a}(1,1,1)$, and
where $v_\parallel$ ($v_\perp$) stands for the velocity parallel (perpendicular) to $\bL$. We use the notation of primed variables refer to the local L point coordinate system.
The bulk gap is the energy difference between
 the $L_{6\pm}$ band edge states which are of opposite parities:  $2\Delta_0= \epsilon_{L_{6-}}-\epsilon_{L_{6+}}$, and is negative reflecting the band inversion in SnTe.

The microscopic basis for this Dirac equation is given by the gap edge wavefunctions

\begin{eqnarray}
\ket{\phi_2^-} & = & -\sin\frac{\theta_-}{2}\ket{+\downarrow} + \cos\frac{\theta_-}{2}\ket{0\uparrow} \\
\ket{T\phi_2^-} & = & -\sin\frac{\theta_-}{2}\ket{-\uparrow} + \cos\frac{\theta_-}{2}\ket{0\downarrow} \\
\ket{\phi_1^+} & = & \cos\frac{\theta_+}{2}\ket{+\downarrow} + \sin\frac{\theta_+}{2}\ket{0\uparrow} \\
\ket{T\phi_1^+} & = & \cos\frac{\theta_+}{2}\ket{-\uparrow} + \sin\frac{\theta_+}{2}\ket{0\downarrow}
\end{eqnarray}
in which $\ket{\phi_2^-}$ and $\ket{T\phi_2^-}$ are the degenerate Kramers pair of the valence band, and $\ket{\phi_1^+}$ and $\ket{T\phi_1^+}$ the degenerate Kramers pair of the conduction band. In these expressions the kets $\ket{m,\sigma}$ are combined spherical harmonic and spin functions; as we have a $p$-band system $m=-1,0,+1$.
The angles $\theta_\pm$ determine the spin mixing in the positive and negative parity gap functions, and reflect the relative strength of the crystal field and spin orbit coupling. For a careful derivation of these gap edge wavefunctions we refer the reader to Ref.~\cite{vol83}.

\section{Derivation of the surface state}

\subsection{Interface equation}
The interface is defined by two $z$ dependent fields, a gap inversion field

\begin{equation}
\Delta = \Delta_0 f(z)
\end{equation}
which enters as $\tau_z$ field in Eq.~\eqref{eq1}
and a band bending field
\begin{equation}
\varphi = \varphi_0 f(z)
\end{equation}
that enters as an additional scalar field.
In these expressions $f(z) \to +F$ as $z \to -\infty$ with $F$ a large positive number defining the ``vacuum gap" and $f(z) \to 1$ as $z \to +\infty$ ($z$ is the coordinate normal to the surface with $z>0$ the material side; unprimed variables will refer to the surface normal coordinate system).
Note that it is necessary to take the same $f(z)$ for the $z$ dependence of both $\Delta(z)$ and $\varphi(z)$, justified in the original context of gap inversion semi-conductor hetero-junctions by a common dependence on alloy composition, for example in band-inverting system Pb$_{1-x}$Sn$_x$Te. As the surface state will turn out to be independent of the particular form of $f(z)$, the assumed common form does not represent a serious impediment to applying this approach also to the material-vacuum interface.

To solve an inhomogeneous Dirac problem the bulk Dirac equation must be rotated from the local L point coordinate system to the surface normal coordinate system. This is implemented via a rotation about the local $z'$-axis, followed by a rotation about the new $y$-axis: $\bR = \bR_y(\beta)\bR_z(\alpha)$, giving for the composite rotation matrix

\begin{equation}
 \bR=\begin{pmatrix}
      \cos\alpha\cos\beta & -\sin\alpha\cos\beta & \sin\beta \\
      \sin\alpha          &  \cos\alpha          & 0           \\
     -\cos\alpha\sin\beta & \sin\alpha\sin\beta  & \cos\beta \\
     \end{pmatrix}
     \label{rotmat}
\end{equation}

To transform the Dirac equation we note
\begin{eqnarray}
  \bsig\bv'\bp'&=&(\bsig\bR^\dagger)(\bR \bv' \bR^\dagger)(\bR   \bp')\nonumber\\
  &=& U^\dagger\bsig U\bv\bp
  \label{10}
\end{eqnarray}
with $\bp=\bR\bp'$ the momentum operators in the surface normal coordinate system
 and where we have used the fact the $U^\dagger \bsig U = \bsig \bR^\dagger$ (for $\bsig$ a row vector) with $U$ the $SU(2)$ form of the rotation matrix $\bR$. This implies that rotation of the bulk coordinate system manifests both as a rotation of the momentum operators and, importantly, as a rotation $U$ acting on the pseudospin space of the Kramers conjugate pair of gap edges states. Writing

\begin{equation}
 \bR\bv \bR^\dagger = \bR\left( v_\perp \mathbbm{1}_3 +            (v_\parallel-v_\perp)
 \begin{pmatrix}
  0 & & \\
  & 0 & \\
  & & 1
 \end{pmatrix}
 \right) \bR^\dagger
\end{equation}
we see that

\begin{equation}
 \bsig\bv\bp = v_\perp \bsig.\bp + (v_\parallel-v_\perp)\bsig.     \bR_z\bp.\bR_z
\end{equation}
where $\bR_z = (R_{xz},R_{yz},R_{zz}) = (\sin\beta,0,\cos\beta)$   and involves only the physically relevant angle $\beta$ (the angle between the local L point z'-axis and the surface normal). This can be expressed as

\begin{equation}
 \bsig\bv\bp = \bsig.\bA + \bsig.\bB p_z
\end{equation}
where with the aid of the vector $\bR_z$ we find

\begin{equation}
 \bsig.\bB = \begin{pmatrix}
                  v_z & v_x \\
                  v_x & -v_z
                 \end{pmatrix}
\end{equation}
with

\begin{eqnarray}
 v_x & = & (v_\parallel - v_\perp)\cos\beta\sin\beta \label{vx} \\
 v_z & = & v_\parallel \cos^2\beta + v_\perp \sin^2\beta \label{vz}
\end{eqnarray}
and
\begin{equation}
 \bsig.\bA = \begin{pmatrix}
                  v_xp_x & (v_\perp + v_\parallel - v_z)p_x + i    v_\perp p_y \\
                  (v_\perp + v_\parallel - v_z)p_x - i v_\perp p_y & -v_xp_x
                 \end{pmatrix}
                 \label{A}
\end{equation}
Altogether this gives an interface Dirac equation

\begin{equation}
H =
\begin{pmatrix}
  \varphi(z) + \Delta(z) & \bsig.\bA +\bsig.\bB p_z \\
  \bsig.\bA +\bsig.\bB p_z  & \varphi(z) - \Delta(z)
 \end{pmatrix}
\label{intD}
\end{equation}

\subsection{Solution of the interface equation}

Following Refs.~\cite{vol85,pan87,pan89} we first square the interface equation and then manipulate it such that it can be factorized into supersymmetry form. The square of the interface equation gives

\begin{equation}
 \Bigg[\Delta^2-\varphi^2-\epsilon^2+                     2\epsilon\varphi + A^2 + B^2 p_z^2  + \{\bsig.\bA,\bsig.        \bB\}p_z 
 +\left(p_z f(z)\right) \begin{pmatrix} 0 & \varphi_0\!-\!\Delta_0 \\
           \varphi_0\!+\!\Delta_0 & 0
          \end{pmatrix}
 \otimes \,\bsig.\bB\Bigg]\Psi = 0
 \label{Hsq}
\end{equation}
which is diagonal except for the last term, which we diagonalize using the following transformation

\begin{equation}
 Z^{-1} \begin{pmatrix} 0 & \varphi_0-\Delta_0 \\
           \varphi_0 + \Delta_0 & 0
          \end{pmatrix}\otimes \bsig.\bB\, Z = v\tau_z\otimes\sigma_z
           \sqrt{\varphi_0^2-\Delta_0^2}
\end{equation}
where

\begin{equation}
 Z = S\otimes U_2 = \frac{1}{\sqrt{2}}
\begin{pmatrix}
 s_- & s_-\\
 s_+ & -s_+
\end{pmatrix} \otimes\frac{1}{\sqrt{2}} \begin{pmatrix}
 u_+ & -u_-\\
 u_- & u_+
\end{pmatrix}
 \label{t6}
\end{equation}
In this transformation the matrix $S$ diagonalizes the field term (i.e. the term involving the band bending and gap inversion fields $\varphi_0$ and $\Delta_0$ respectively)
and has the $s_\pm$ given by
\begin{equation}
s_\pm =                                                            \frac{\sqrt{\varphi_0\pm\Delta_0}}{\sqrt{\left|\Delta_0\right|}}
\end{equation}
while the $U_2$ part diagonalizes the matrix $U_2^\dagger \bsig.\bB U_2 = v \sigma_z$, 
and has

\begin{equation}
 u_\pm= \sqrt{1\pm v_z/v}
\end{equation}
and where $v = \sqrt{v_x^2+v_z^2}$ which with help of Eq.~\eqref{vx}, \eqref{vz} can be written as

\begin{equation}
 v = \sqrt{v_\parallel^2 \cos^2\beta + v_\perp^2 \sin^2\beta}
 \label{v}
\end{equation}

In order to factorize Eq.~\eqref{Hsq} we must eliminate the linear term in $p_z$, which we do by the unitary transformation

\begin{equation}
 U_1 = \mathbbm{1}_4 e^{i\kappa z}
 \label{t7}
\end{equation}
which will, by the action of the $p_z^2$ operator in Eq.~\ref{Hsq}, introduce a new linear in $p_z$ term to the equation.  We can then choose $\kappa$, a free parameter, such that this exactly cancels the original linear in $p_z$ term. This requires

\begin{equation}
 \kappa =\frac{(v_\parallel^2-v_\perp^2)\sin\beta\cos\beta        p_x}{v^2}
\end{equation}
Finally by introducing the superpotential

\begin{equation}
W(z) = \sqrt{\Delta_0^2-\varphi_0^2}\left(f(z) +                   \frac{\epsilon\varphi_0}{\Delta_0^2-\varphi_0^2}\right)
\end{equation}
all field terms can be expressed through $W(z)$ and Eq.~\eqref{Hsq} factorizes in a sypersymmetric form (as a product of  creation and annihilation operators) to give

\begin{equation}
(W(z) + i v \tau_z\otimes\sigma_z p_z)(W(z) - i v\tau_z\otimes\sigma_z p_z)Z^{-1}\Psi =  \left(\frac{\Delta_0^2\epsilon^2}{\Delta_0^2-\varphi_0^2}
-\frac{v_\perp^2 v_\parallel^2}{v^2}p_x^2-v_\perp^2p_y^2
\right)Z^{-1}\Psi
 \label{SS}
\end{equation}
where terms involving $\bsig.\bA$ and $\bsig.\bB$ have been evaluated explicitly.
A zero mode solution of the squared interface equation is thus obtained by the action of the annihilation operator

\begin{equation}
 (W(z) - i v \tau_z\otimes\sigma_z p_z)\ket{\pm} = 0
\end{equation}
from which we have two independent solutions

\begin{eqnarray}
 \ket{+} & = & \ket{\uparrow} \otimes \ket{\uparrow} \phi(z) \\
 \ket{-} & = & \ket{\downarrow} \otimes \ket{\downarrow} \phi(z)
\end{eqnarray}
with

\begin{equation}
\phi(z) = e^{\frac{1}{\hbar v}\int_0^zdz'\,W(z')}
\label{phi}
\end{equation}
For these to be normalizable we require that the superpotential be real valued and change sign asymptotically. This implies not only gap inversion but also

\begin{equation}
 |\varphi_0| < |\Delta_0|
 \end{equation}
and furthermore gives a ``continuum energy'' at which the surface state smoothly joins the bulk band and ceases to exist:

\begin{equation}
 \epsilon_c =  -\frac{\Delta_0^2-\varphi_0^2}{\varphi_0}
\end{equation}
Given these conditions setting the right hand side of Eq.~\eqref{SS} to zero then yields the  spectrum of the surface state

\begin{equation}
 \epsilon = \pm \gamma \sqrt{\Biggl(\frac{v_\parallel              v_\perp}{v}\Biggr)^2 p_x^2 + (v_\perp)^2 p_y^2}
\end{equation}
where

\begin{equation}
 \gamma = \sqrt{1-\frac{\varphi_0^2}{\Delta_0^2}}
\end{equation}
is a velocity reduction factor. As the zero mode solutions of the squared interface equation are degenerate, we require the linear combination $c_+\ket{+}+c_-\ket{-}$ that gives an eigenstate of the unsquared interface equation. This can be obtained by diagonalizing $U_1^\dagger Z^{-1} (H_{D}+\varphi) Z U_1$ in the sub-space $\ket{\pm}$. Transforming the interface equation we find

\begin{equation}
 U_1^\dagger Z^{-1} (H_{D}+\varphi) Z U_1 =
 \begin{pmatrix}
  \varphi(z) + \frac{\varphi_0}{\sqrt{\varphi_0^2-\Delta_0^2}}     U^\dagger\bsig\bV\bp U &
  \Delta(z) - \frac{\Delta_0}{\sqrt{\varphi_0^2-\Delta_0^2}}       U^\dagger\bsig\bV\bp U \\
  \Delta(z) + \frac{\Delta_0}{\sqrt{\varphi_0^2-\Delta_0^2}}       U^\dagger\bsig\bV\bp U &
  \varphi(z) - \frac{\varphi_0}{\sqrt{\varphi_0^2-\Delta_0^2}}     U^\dagger\bsig\bV\bp U
 \end{pmatrix}
\end{equation}
where $U = U_2 U_1$, and for the effective Hamiltonian yielding $c_\pm$

\begin{equation}
 H_{eff}=\begin{pmatrix}
          \bra{+}Z^{-1}(H_D + \varphi) Z \ket{+}  &  \bra{+}Z^{-   1}(H_D + \varphi) Z \ket{-}\\
          \bra{-}Z^{-1}(H_D + \varphi) Z \ket{+}  &  \bra{-}Z^{-   1}(H_D + \varphi) Z \ket{-}
         \end{pmatrix}
         \label{t2}
\end{equation}
we then find

\begin{equation}
 H_{eff}=
 \begin{pmatrix}
  0 &
  i\gamma\left(\frac{v_\parallel v_\perp}{v} p_x - i v_\perp       p_y\right) \\
   -i\gamma\left(\frac{v_\parallel v_\perp}{v} p_x + i v_\perp     p_y\right) &
  0
 \end{pmatrix}
\end{equation}
which is the surface state Dirac-Weyl effective Hamiltonian describing the spectrum obtained by setting the right hand side of Eq.~\eqref{SS} to zero. The coefficients $c_\pm$ are then just the components of the DW eigenfunction
\begin{equation}
 \psi_{DW}=\begin{pmatrix} c_+ \\ c_- \end{pmatrix} =              \frac{1}{\sqrt{2}}\begin{pmatrix}
                         e^{-i(\frac{\theta_{\bk}}{2}-\frac{\pi}{4})} \\
                         l e^{i(\frac{\theta_{\bk}}{2}-\frac{\pi}{4})}
                        \end{pmatrix}
e^{i\bk.\brr}
\label{DWwf}
\end{equation}
where $l=\pm 1$ labels the electron and hole cones and the angle   $\theta_{\bk}$ is given by

\begin{equation}
 \theta_{\bk}= \tan^{-1}\Bigg(\frac{v p_y}{v_\parallel p_x}        \Bigg)
 \label{tk}
\end{equation}

We thus have the surface state solution of $Z^{-1}(H_D+\varphi) Z$:

\begin{equation}
 \psi=\frac{1}{\sqrt{2}}\begin{pmatrix}
       e^{-i(\frac{\theta_{\bk}}{2}-\frac{\pi}{4})} \\ 0 \\ 0 \\ l                 e^{i(\frac{\theta_{\bk}}{2}-\frac{\pi}{4})}
      \end{pmatrix}
      e^{i\bk.\brr}\phi(z)
\end{equation}
and by acting on this with $U_1 Z$ we finally find the surface state eigenfunction of the interface equation Eq.~\eqref{intD}:

\begin{equation}
 \Psi = \Biggl[\frac{e^{-                                          i(\frac{\theta_{\bk}}{2}-\frac{\pi}{4})}}{2\sqrt{2}}\begin{pmatrix}
                            s_- u_+ \\ s_-u_- \\ s_+ u_+ \\ s_+ u_-
                            \end{pmatrix}
              +\frac{l                                             e^{i(\frac{\theta_{\bk}}{2}-\frac{\pi}{4})}}{2\sqrt{2}}\begin{pmatrix}
                            -s_- u_- \\ s_-u_+ \\ s_+ u_- \\ -s_+  u_+
                           \end{pmatrix}\Biggr] e^{\frac{1}{\hbar  v}\int_0^zdz'\,W(z')} e^{i(\kappa z+\bk.\brr)}
                           \label{intS}
\end{equation}
the $\kappa$ phase represents a mixing of the propagating and exponential solutions   that occurs once the surface normal is not aligned with the bulk L point coordinate system, and where $s_\pm$ and $u_\pm$ arise from the transformation $Z$ (see Eq.~\ref{t6}-\ref{t7}).

\subsection{The spin texture}

The solution to the interface equation, Eq.~\eqref{intS}, represents the surface state in the basis of bulk band edge states. To calculate spin texture we require the surface state in terms of the microscopic variables of the bulk insulator: spin and angular momentum. However, as the interface Dirac equation was solved in the surface normal coordinate system, and our original basis functions expressed in the bulk L coordinate system, we must transform the interface wavefunction back to the local L coordinate system.
This requires undoing the transformation Eq.~\eqref{10} by

\begin{figure}
\centering
\includegraphics[width=0.35\textwidth]{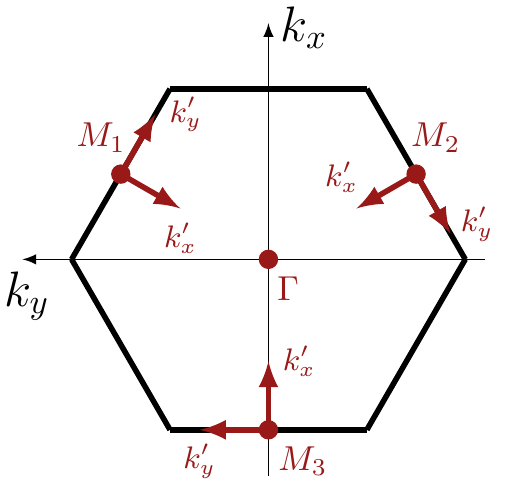}
\caption{Surface Brillouin zone of the (111) facet of SnTe, showing the 4 high symmetry points at which Dirac-Weyl topological surface states are found: $\Gamma$ and $M_1$-$M_3$. For the latter the local coordinate system of the spin texture is also shown.}
\label{fig1}
\end{figure}

\begin{equation}
 \Psi' = \begin{pmatrix} U^\dagger & 0 \\ 0 & U^\dagger            \end{pmatrix} \Psi
\end{equation}
where $U$ is the SU(2) rotation corresponding to $R$ given in Eq.~\eqref{rotmat} given by

\begin{equation}
U = \begin{pmatrix}
e^{-i\alpha/2} \cos\frac{\beta}{2} & -e^{i\alpha/2} \sin\frac{\beta}{2} \\
e^{-i\alpha/2} \sin\frac{\beta}{2}
&
e^{i\alpha/2} \cos\frac{\beta}{2}
\end{pmatrix}
\end{equation}
This sends the $u_\pm$ coefficients to $u'_\pm$ coefficients as

\begin{eqnarray}
 u'_+ & = & \cos\frac{\beta}{2} u_+ + \sin\frac{\beta}{2} u_- \\
 u'_- & = & -\sin\frac{\beta}{2} u_+ + \cos\frac{\beta}{2} u_-
 \label{t11}
\end{eqnarray}
giving for the surface state wavefunction in microscopic variables

\begin{eqnarray}
 \Psi' = \Bigg[
 \frac{c_+}{2}\Biggl(
 e^{i\alpha/2}s_-u'_+\ket{\phi_2^-} + e^{-i\alpha/2}s_-u'_-        \ket{T\phi_2^-} + e^{i\alpha/2}s_+u'_+\ket{\phi_1^+} \nonumber \\
 + e^{-i\alpha/2}s_+u'_-\ket{T\phi_1^+}
 \Biggr) + \frac{c_-}{2}\Biggl(
 -e^{i\alpha/2}s_-u'_-\ket{\phi_2^-} + e^{-i\alpha/2}s_-u'_+       \ket{T\phi_2^-} + \nonumber \\
 e^{i\alpha/2}s_+u'_-\ket{\phi_1^+} - e^{-i\alpha/2}s_+u'_+\ket{T\phi_1^+}
 \Biggr) \Bigg] \phi(z) e^{i\kappa z}
\label{t10}
\end{eqnarray}
where the coefficients $c_\pm$ are given in Eq.~\eqref{DWwf}.
This has the form
$\ket{\Psi'} = \ket{X'} + l\ket{TX'}$  from which the spin texture $\bmm'_\bk = \mel{\Psi'}{\bsig}{\Psi'}$ is evaluated as

\begin{eqnarray}
 \bmm'_{\bk} =\nonumber\\
  \frac{l}{2} \Biggl(1 +                             \frac{\varphi_0}{\Delta_0}\Biggr)
                \Biggl(
                -\frac{v_\perp}{v}\sin\beta\sin\theta_{\bk}        \bra{\phi_1^+}\bsig\ket{\phi_1^+} -                                \frac{v_\parallel}{v}\cos\beta\sin\theta_{\bk} \text{Re}\,         e^{i\alpha} \bra{T\phi_1^+}\bsig\ket{\phi_1^+}\nonumber\\
 + \cos\theta_{\bk} \text{Im}\,e^{i\alpha} \bra{T\phi_1^+        }\bsig\ket{\phi_1^+}
 \Biggr)\nonumber\\
  +\frac{l}{2} \Biggl(1 - \frac{\varphi_0}{\Delta_0}\Biggr)
 \Biggl(
 \frac{v_\perp}{v}\sin\beta\sin\theta_{\bk} \bra{\phi_2^-          }\bsig\ket{\phi_2^-} +                                             \frac{v_\parallel}{v}\cos\beta\sin\theta_{\bk} \text{Re}\,         e^{i\alpha} \bra{T\phi_2^-}\bsig\ket{\phi_2^-} \nonumber\\
 - \cos\theta_{\bk} \text{Im} \,e^{i\alpha}\bra{T\phi_2^-        }\bsig\ket{\phi_2^-}
 \Biggr) \nonumber \\
  l\sqrt{1-                                                    \frac{\varphi_0^2}{\Delta_0^2}}\Biggl(\frac{v_\perp}{v}\sin\beta\cos\theta_{\bk}\text{Re}\mel{\phi_1^+}{\bsig}{\phi_2^-}
 -\frac{v_\parallel}{v}\cos\beta \cos\theta_{\bk}\text{Re}\,       e^{i\alpha}\mel{T\phi_1^+}{\bsig}{\phi_2^-}  \nonumber \\
  +\sin\theta_{\bk}\text{Im}\,e^{i\alpha}\mel{T\phi_1^+          }{\bsig}{\phi_2^-} \Biggl)
 \label{sptex}
\end{eqnarray}
where the angle $\theta_k$ is defined in Eq.~\eqref{tk}.
We see that the surface band bending weights the contributions to the texture from the conduction band, valence band, and conduction-valence coupling through the factors $(1+\varphi_0/\Delta_0)/2$, $(1-\varphi_0/\Delta_0)/2$, and $\sqrt{1-                                                    \frac{\varphi_0^2}{\Delta_0^2}}$ respectively, with the corresponding terms involving the microscopic degrees of freedom of the bulk through the matrix elements of the vector of Pauli matrices $\bsig$. As shown in the previous section the band bending cannot bend bands outside the bulk gap $|\varphi_0| < |\Delta_0|$ without destroying the surface state. Tuning the band bending $\varphi_0$ through the gap the weight factors describe a continuous shift from a dominant conduction band contribution for upward band bending to a dominant valence band contribution for downward band bending. The inter-band term contributes maximally for zero band bending, falling to zero for the limits $\varphi_0=\pm |\Delta_0|$.

Although the expressions Eq.~\eqref{intS}-\eqref{sptex} have been derived using notation relevant to the weak topological insulator SnTe, they are general expressions representing the solution for the surface state of any bulk Dirac manifold whose principal axis are misoriented from the surface normal. The spin texture Eq.~\eqref{sptex} is thus a general expression representing the entwining of the bulk microscopic variables and surface banding bending.

\section{Hyperbolic spin textures at the M point Dirac-Weyl cone}
\begin{figure}
\centering
\includegraphics[width=0.7\textwidth]{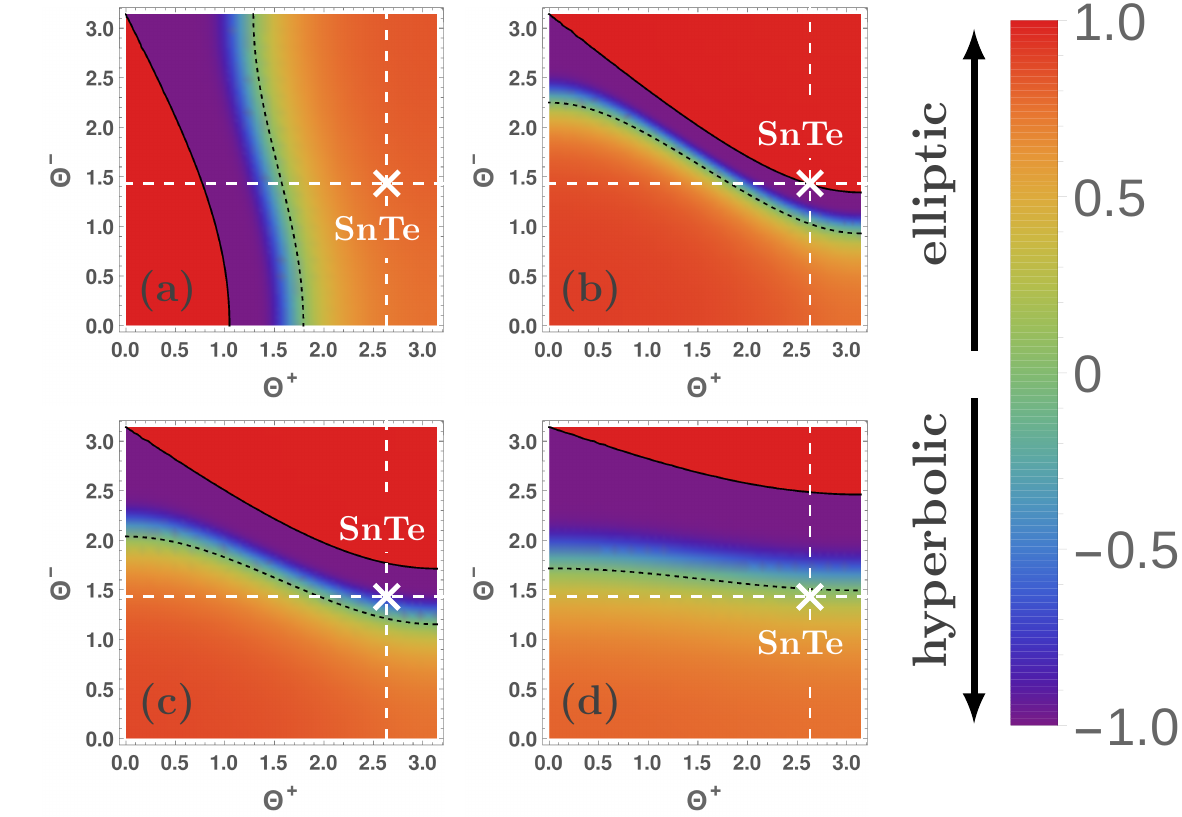}
\caption{{\it Dependence of surface spin texture on the balance of spin orbit and crystal field strength in the bulk}. The topology of the spin texture can be characterized by a number $\eta = \tanh\frac{m_y}{m_x}$, positive for helical and negative for hyperbolic textures. Shown is the dependence of $\eta$ on the spin mixing angles $\theta_\pm$, which encode the balance of spin-obit interaction and crystal field in the bulk.  Each panel represents a different value the surface band bending parameter: $\varphi_0/|\Delta_0|$ = -0.60, 0.24, 0.40, 0.80 (panels (a) to (d) respectively). }
\label{fig2}
\end{figure}

\begin{table}
 \begin{tabular}{cc}
  Matrix element & Polarization vector \\ \hline
  $\mel{\phi_1^+}{\bsig}{\phi_1^+}$  & $-\cos\theta_+(0,0,1)$ \\
  $\mel{\phi_2^-}{\bsig}{\phi_2^-}$  & $+\cos\theta_-(0,0,1)$ \\
  $\mel{T\phi_1^+}{\bsig}{\phi_1^+}$ & $\sin^2\frac{\theta_+}{2}(1,i,0)$ \\
  $\mel{T\phi_2^-}{\bsig}{\phi_2^-}$ & $\cos^2\frac{\theta_-}{2}(1,i,0)$ \\
 \end{tabular}
 \caption{Matrix elements of the band edge states in the (111) coordinate system.}
 \label{matel}
\end{table}

We now consider the spin texture at both the $\Gamma$ and M points of (111) surface of SnTe (see Fig.~\ref{fig1}). In experiment these are separated in energy by $\sim 170$~meV\cite{tan13} and so are, in principle, individually accessible and, more importantly, do not couple (as the surface states do on the (100) facet\cite{shi14}). Insertion of $\alpha=\beta=0$ into Eq.~\eqref{sptex} along with the use of Table \ref{matel} results in a helical texture $\bmm_\bk = l \rho_\perp (\sin\theta_{\bk},-\cos\theta_{\bk})$ where

\begin{equation}
\rho_\perp = 
-\frac{1}{2}\left(1+                                              \frac{\varphi_0}{\Delta_0}\right)\sin^2\frac{\theta_+}{2} + 
\frac{1}{2}\left(1-                                               \frac{\varphi_0}{\Delta_0}\right)\cos^2\frac{\theta_-}{2}
\label{50}
\end{equation}
and the angle $\theta_{\bk}$ is defined in Eq.~\eqref{tk} and $l=\pm 1$ refers to the electron (+1) or hole (-1) cones.
No other textures are possible at the $\Gamma$ point. The winding number of the helical texture depends on the band bending, reflecting the fact that the conduction and valence band contribute with opposite sign of the winding number to the texture, and so by tuning the band bending to move from a regime of dominant conduction to dominant valence spin texture, once can thus change the texture winding number.

For the M points, the angle 
$\beta=\cos^{-1}\frac{1}{3}$, and the spin texture can again be determined from Eq.~\eqref{sptex} and Table \ref{matel}. Now, however, as the texture is expressed in terms of the bulk band edge functions expressed in the local L point coordinate system, we require rotation back to the surface normal frame. The $R_z(\alpha)$ rotation is not physically significant, and this leaves the $R_y(\beta)$ rotation to be performed

\begin{equation}
 \bmm_{\bk} = l \begin{pmatrix}
      \cos\beta & 0 & \sin\beta \\
      0          &  1          & 0           \\
     -\sin\beta & 0  & \cos\beta \\
     \end{pmatrix}
     \begin{pmatrix}
      \rho_\perp\frac{v_\parallel}{v}\cos\beta\sin\theta_{\bk} \\
      -\rho_\perp \cos\theta_{\bk} \\
      \rho_\parallel\frac{v_\perp}{v}\sin\beta \sin\theta_{\bk}
     \end{pmatrix}
\end{equation}
where

\begin{equation}
\rho_\parallel = \frac{1}{2}\left(1+                              \frac{\varphi_0}{\Delta_0}\right)\cos\theta_+ + 
\frac{1}{2}\left(1-\frac{\varphi_0}{\Delta_0}\right) \cos\theta_-
\label{51}
\end{equation}

The right hand side of this equation represents the spin polarization in the local L frame, which already has a non-zero $z'$ component. This is different from the result in Ref.~\cite{zha12} (where in the local frame the $z'$ component of spin is identically zero) and has an important consequence for the M point spin texture. 

As the constants $\rho_\perp$ and $\rho_\parallel$ 
can evidently be both positive or negative quantities, upon rotating back to the surface normal frame 
\begin{equation}
\bmm_{\bk} =
l
\begin{pmatrix}
 m_x\sin\theta_{\bk}\\
 -m_y \cos\theta_{\bk}\\
 m_z \sin\theta_{\bk}
\end{pmatrix}
\label{spin-pol}
\end{equation}
the $\sin\beta$ and $\cos\beta$ combinations of  $\rho_\perp$ and $\rho_\parallel$ 
can give the spin texture components

\begin{equation}  
m_x=\rho_\parallel \Biggl(\frac{v_\perp}{v}\Biggr)\sin^2\beta+       \rho_\perp \Biggl(\frac{v_\parallel}{v}\Biggr)\cos^2\beta
\end{equation}
and

\begin{equation}
m_y=\rho_\perp
\end{equation}
of either sign. This freedom of the sign in $m_x$ and $m_y$ allows for both helical and hyperbolic spin textures to occur (if both have the same sign the texture is helical, and if they have different signs the texture is hyperbolic). Note that if the $z'$ polarization in the local L frame was zero the only possible surface spin texture would be helical. Finally, we find the out-of-plane magnetization in the surface normal frame to be

\begin{equation}
m_z                                                                =\frac{\sin(2\beta)}{2}\Biggl[\Biggl(\frac{v_\perp}{v}\Biggr)\rho_\parallel-\Biggl(\frac{v_\parallel}{v}\Biggr)\rho_\perp\Biggr]
\end{equation}
which is generally non-zero for an misoriented facet, in agreement with Ref.~\cite{zha12}.

\begin{figure}
\centering
\includegraphics[width=0.9\textwidth]{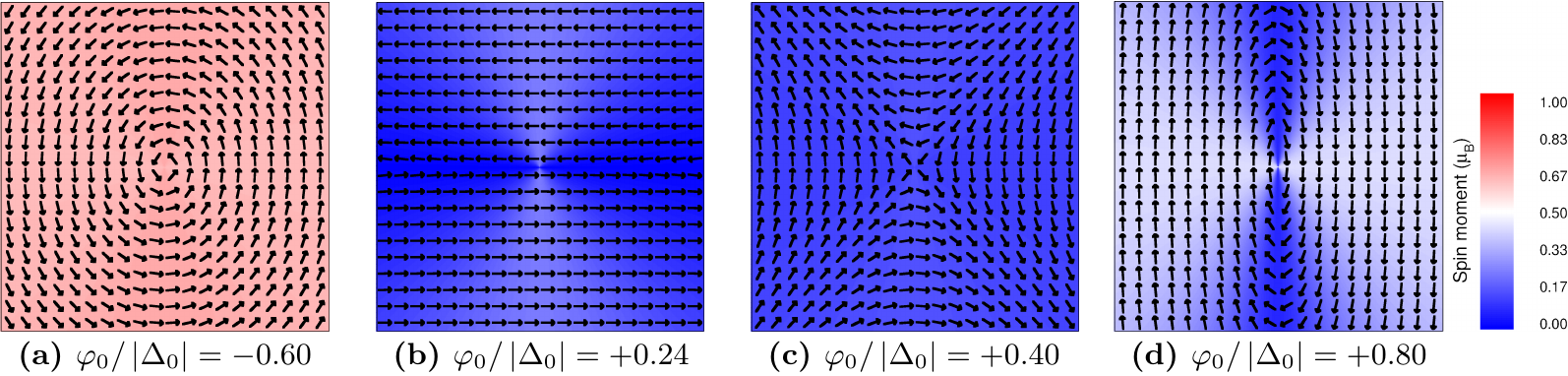}
\caption{{\it Helical and hyperbolic spin texture at the M point Dirac-Weyl cone of the (111) surface of SnTe}. Shown are the spin textures calculated for values of the band bending parameter, $\varphi_0/|\Delta_0|$ = -0.60, 0.24, 0.40, 0.80 (panels (a) to (d) respectively). Note that the downward band bending of the Sn terminated surface typically seen in experiment corresponds to a positive value of $\varphi_0/|\Delta_0|$.}
\label{fig3}
\end{figure}

The expressions for $\rho_\perp$ and $\rho_\parallel$ involve both band bending and the spin mixing angles of the bulk gap edge wavefunction. To explore how these impact the spin texture we define a ``texture parameter" $\tanh \frac{m_y}{m_x}$, which is positive for a helical texture and negative for a hyperbolic texture, and in Fig.~\ref{fig2} we show how this parameter depends on both spin mixing and band bending. Evidently, the dependence is rich, showing that the spin structure of the surface state is strikingly non-universal and material dependent. In each of these plots the spin mixing angles corresponding to bulk SnTe (taken from Ref.~\cite{vol83}) are indicated, and in Fig.~\ref{fig3} we plot the corresponding spin texture. For the downward band-bending seen in experiment\cite{tan13,tas14} for the Sn terminated surface we find a winding number of -1 and an elliptically distorted texture, in agreement with {\it ab-initio} results\cite{shi14}. However, our calculations reveal that by tuning the band bending, panels (a) to (c), a rich evolution of the spin texture is predicted.

\section{Discussion}

We have shown that misorientation between the principle axis of an anisotropic low energy Dirac manifold and the surface normal of a topological insulator unlocks a rich variety of surface spin textures. Underpinning this is the fact that the miorientation requires rotation within the degenerate Kramers sub-space of the band gap edge states. The surface state is sensitive to band bending, which determines the relative weight of the bulk valence and conduction gap edge states in the surface state. Shifting of the bulk spectrum by a band bending energy $\varphi_0$ is allowed for energy shifts within the bulk gap $\Delta_0$: $-|\Delta_0| < \varphi_0 < +|\Delta_0|$ with, at these limits, the surface state pure valence and pure conduction respectively (outside this energy range the surface state is destroyed). This weighting of conduction and valence contributions allows tuning of the spin texture between helical and hyperbolic by band bending.
The nature of the surface spin structure turns out, in addition, to depend sensitively on the spin-mixing within the bulk band edge states, and alteration of this can also tune between helical and hyperbolic spin textures. The surface spin texture of a TI for the case of misorientation is thus strikingly non-universal, dependent both on microscopic physics of the gap edge wavefunctions through the balance of spin-orbit to crystal field in the bulk, and the band bending at the interface to the vacuum.

For the case of the (111) facet of SnTe, a downward band bending generally results in a distorted helical texture, with only a narrow region of the gap resulting in a hyperbolic texture. In agreement with {\it ab-initio} work\cite{shi14} we find for the M point an out-of-plane texture and a winding number of -1 for the conduction band texture. As the extension of the surface state is over several lattice constants it should be possible by doping of the polar (111) facet to tune band bending, resulting in a degree of control over spin texture.

\section*{References}
%\bibliography{literature}

\providecommand{\newblock}{}

\end{document}